\documentclass[12pt]{article}
\usepackage{epsf}
\begin{document}
\def\muc{$\mu$C}
\def\be{\begin{equation}}
\def\ee{\end{equation}}
\def\bea{\begin{eqnarray}}
\def\eea{\end{eqnarray}}
\def\ba{\begin{array}}
\def\ea{\end{array}}
\def\bra#1{\langle #1 |}
\def\ket#1{| #1\rangle }
\def\esp#1{e^{#1}}
\def\slash#1{\setbox0=\hbox{$#1$}#1\hskip-\wd0\dimen0=5pt\advance
       \dimen0 by-\ht0\advance\dimen0 by\dp0\lower0.5\dimen0\hbox
         to\wd0{\hss\sl$\diagup$\hss}}
\def\mm{\mu^+\mu^-}
\def\anti{\overline}
\def\ha{A^0}
\def\hl{h^0}
\def\hh{H^0}
\def\ds{\displaystyle}
\def\gamh{\Gamma_H}
\def\eb{E_{\rm beam}}
\def\deb{\Delta E_{\rm beam}}
\def\sigm{\sigma_M}
\def\sigmmax{\sigma_M^{\rm max}}
\def\sigmmin{\sigma_M^{\rm min}}
\def\sige{\sigma_E}
\def\dsigm{\Delta\sigma_M}
\def\mh{M_H}
\def\lyear{L_{\rm year}}

\def\wstar{W^\star}
\def\zstar{Z^\star}
\def\ie{{\it i.e.}}
\def\etal{{\it et al.}}
\def\eg{{\it e.g.}}
\def\pzero{P^0}
\def\mt{m_t}
\def\mpzero{M_{\pzero}}
\def\mev{~{\rm MeV}}
\def\gev{~{\rm GeV}}
\def\gam{\gamma}
\def\lsim{\mathrel{\raise.3ex\hbox{$<$\kern-.75em\lower1ex\hbox{$\sim$}}}}
\def\gsim{\mathrel{\raise.3ex\hbox{$>$\kern-.75em\lower1ex\hbox{$\sim$}}}}
\def\ntc{N_{TC}}
\def\epem{e^+e^-}
\def\tauptaum{\tau^+\tau^-}
\def\lplm{\ell^+\ell^-}
\def\anti{\overline}
\def\mw{M_W}
\def\mz{M_Z}
\def\fbi{~{\rm fb}^{-1}}
\def\mupmum{\mu^+\mu^-}
\def\rts{\sqrt s}
\def\sigrts{\sigma_{\tiny\rts}^{}}
\def\sigrtssq{\sigma_{\tiny\rts}^2}
\def\sigrtsprime{\sigma_{E}}
\def\nsigrts{n_{\sigrts}}
\def\gampzero{\Gamma_{\pzero}}
\def\pzerop{P^{0\,\prime}}
\def\mpzerop{M_{\pzerop}}

\begin{center}
{\Large\bf\boldmath {Precision Measurements at a Muon Collider }}
\\ \rm \vskip1pc {\large
R. Casalbuoni }\vskip0.2cm{\it{Dipartimento di Fisica,
Universit\`a di Firenze,  and I.N.F.N., Sezione di Firenze,
I-50125 Firenze, Italia}}
\end{center}

%
This talk is mainly based on the CERN Workshop held at CERN in
1998 \cite{CERN}. I will not cover all the possible physics
aspects of a muon collider (\muc) at various energies, but I will
rather concentrate on the most peculiar aspects of such a
machine. A {\muc} has all the advantages of a lepton vs. a
hadronic collider, in particular the fact that the effective
energy is bigger. But a muon machine has also big advantages
compared with an electron-positron collider. Namely, muons have
negligible synchrotron radiation giving the possibility to make
muon circular colliders of a relatively modest size. Also muons do
not exhibit beamstrahlung. This allows to reach a very small beam
energy spread, as low as as $3\times 10^{-5}$ times the beam
energy. This will appear in our discussion as the key parameter
for a \muc. At electron colliders it is not so easy to monitor
this parameter. On the other hand, the natural polarization of
muons allows a very good energy determination (at the level of
$10^{-6}$ or better) and also the measurement of the energy
spread. Conventionally the energy spread is parameterized as
$\Delta E_{\rm beam}/E_{\rm beam}=0.01 R(\%)$. It follows that
the center of mass energy spread $\sigma_E$ is given by $
\sigma_E\approx 0.007 R(\%)E$. In principle it is possible to
improve the $R$ factor by reducing the luminosity of the machine.
For a {\muc} of $\sqrt{s}=100~GeV$ we will assume the following
luminosities and related $R$ values \cite{gunion1,CERN}
\begin{center}
\begin{tabular}{|c|c|c|c|}
\hline
            $R(\%)$            & 0.12 & 0.01 & 0.003 \\
            \hline
 $L_{\rm year}({\rm fb}^{-1})$ &  1   & 0.22 &  0.1  \\
 \hline
\end{tabular}
\end{center}
These features make the {\muc} an ideal machine for exploring
thresholds and for studying very narrow resonances. There is
another bonus connected with the flavor of muons. Higgs-like
resonances cannot be produced in the $s$-channel at electron
colliders, since the cross-section is by far too low. However the
cross-section for muons to produce Higgs or particles with
coupling proportional to the lepton mass are $4\times 10^4$
bigger. Therefore one has  the possibility to see directly a
Higgs-like particle (under certain conditions to be specified
later on). In this contribution I will confine my analysis to the
study of narrow resonances. I will assume a Breit-Wigner shape
for the cross-section to produce a spin $j$ resonance in the
$s$-channel with decay $R\to F$
\begin{equation}\label{BW}
\sigma^{F}(E)=4\pi(2 j+1)\frac{\Gamma(R\to\ell^+\ell^-)
\Gamma(R\to F)}{(E^2-M^2)^2+M^2\Gamma^2}
\end{equation}
I will ignore the running of the total width with the energy due
to the very narrow resonance assumption $\Gamma\ll M$. Assuming
for the moment a Gaussian shape for the beam we get the observed
cross-section after convolution with the energy distribution of
the beam $f(E)$. We get
\begin{equation}\label{convolution}
\sigma_c^{F}(E)=4\pi(2j+1) \Gamma(R\to\ell^+\ell^-) \Gamma(R\to
F) h(\Gamma,\sigma_E,E)
\end{equation}
where
\begin{equation}\label{convolution2}
h(\Gamma,\sigma_E,E)=\int \frac{f(E-
E')}{({E'}^2-M^2)^2+M^2\Gamma^2}d E'
\end{equation}
The convoluted production cross-section evaluated at the peak is
given by
\begin{equation} \sigma_c(M)={4\pi B_{\ell^+\ell^-}\over
M^2}\times\left\{\matrix{1 & \Gamma\ll\sigma_M\cr \frac 12
\sqrt{\frac\pi 2} \frac\Gamma{\sigma_M} &
\Gamma\gg\sigma_M}\right\} \label{largegam}
\end{equation}
In the first case $\sigma_c(M)$ gives directly the branching ratio
$B_{\ell^+\ell^-}$. In the other case one has to get $\Gamma$
from a scan of the resonance. Notice that the statistical errors
are worse in the second case, since $\Gamma/\sigma_M\ll 1$.
Furthermore $\sigma_c(M)$ depends on $\sigma_M$, meaning that the
errors on the latter quantity will induce errors on the resonance
parameters. This has been studied in ref. \cite{JHEP}, where it
has been shown that for $\sigma_M\approx \Gamma$ the errors on
the branching ratio and on the width, using a scanning procedure,
are about 3 times $\Delta\sigma_M/\sigma_M$. Therefore an effort
should be done to keep the errors on the energy spread at the
level of per cent. A different method for keeping the
$\sigma_M$-induced errors under control has been discussed in
ref. \cite{ratio}. Of course one could try to optimize the net
statistical plus systematic errors, but in general this requires
a prior knowledge of the width. For very narrow resonances as the
Higgs or a pseugoldstone boson one has to start with the smallest
possible value of $R$. This means to operate with $R\approx
0.003\%$ with a corresponding sacrifice for the luminosity.

{\it Standard-Model like Higgs} - Let me now start  discussing the
$s$-channel production of a SM-like Higgs. This detection mode is
possible only for relatively small Higgs masses up to $140\div
150~GeV$. For higher values of the mass the channels $WW^*$ and
$ZZ^*$ start to open up and the Higgs becomes rapidly a broad
resonance. Correspondingly both the branching ratios $Br(H\to
b\bar b)$  and $Br(H\to\mu^+\mu^-)$ decline, making this
production mode useless. In the range of masses where the
cross-section is a decent one, one assumes that the Higgs has been
seen at LHC and/or at a NLC with a mass uncertainty of about
$100~MeV$. With this information one can center on
$\sqrt{s}\approx M_H$ via scanning. In the previous range of
masses the typical Higgs width is about $1\div 10~MeV$ and, as a
consequence, also running at the minimum value of $R$, we have
$\Gamma_H\approx\sigma_M$. Therefore it is crucial to control the
uncertainty on $\sigma_M$ at the level of per cent. By doing that
one can avoid the contamination of the statistical errors with
the systematics induced by $\sigma_M$. Using the scan procedure
one finds that for a total accumulated luminosity of $L=0.4$
fb$^{-1}$ (corresponding to about 4 years of running) the
statistical errors for the various line shape quantities are given
in the following Table.
\begin{center}
\begin{tabular}{|c|c|c|c|c|c|c|}
\hline
 Quantity & \multicolumn{6}{c|}{Errors for the scan procedure} \\
\hline \hline
 {\bf Mass (GeV)} & {\bf 100} & {\bf 110} & {\bf 120}
 & {\bf 130} & {\bf 140}& {\bf 150}\\
\hline $\sigma_cB(b\anti b)$ & $4\%$ & $3\%$ & $3\%$ & $ 5\%$ & $
9\%$& $ 28\%$ \\
\hline $\sigma_cB(W\wstar)$ & $32\%$ & $15\%$ & $ 10\%$ &
$ 8\%$ & $7\%$ & $9\%$\\
\hline $\sigma_cB(Z\zstar)$ & $-$ & $190\%$ & $50\%$ &
$ 30\%$ & $26\%$ & $34\%$\\
\hline
 {$\Gamma_H$} & {$ 30\%$} & {$ 16\%$} &
 {$ 16\%$}
 & {$18\%$} & {$29\%$} & {$105\%$}\\
\hline
\end{tabular}
\end{center}
The comparison with TESLA possibilities with $500$ fb$^{-1}$/yr
does not look too favorable. In fact as illustrated in the
following Table  \cite{TDR}, in 1 year of running, TESLA can do
better than the {\muc} in 4 years. Although the TESLA TDR does not
include the systematic errors, and a definite comparison should
wait for a careful analysis of the latter ones, the {\muc} needs
to increase its luminosity to be competitive with TESLA on this
particular issue.
\begin{center}
\begin{tabular}{|c|c|c|c|}
\hline
 Quantity & \multicolumn{3}{c|}{Errors at TESLA} \\
\hline \hline
 {\bf Mass (GeV)} & {\bf 120} & {\bf 140} & {\bf 150}
 \\
\hline $B(b\anti b)$ & $2.4\%$ & $2.6\%$ & $6.5\%$ \\
\hline $B(W\wstar)$ & $5.1\%$ & $2.5\%$ & $2.1\%$ \\
\hline $B(Z\zstar)$ & $-$ & $-$ & $16.9\%$ \\
\hline {$\Gamma_h^{\rm
tot}=\frac{\Gamma(WW)=\Gamma(ZZ)}{B(WW^*)}$}
& {$5.6\%$} & {$3.7\%$} & {$3.6\%$}\\
\hline
\end{tabular}
\end{center}
Given this situation, it is clear that also for the possible
implications of these measurements about SM extensions, as SUSY
etc., the {\muc} is not competitive with TESLA. However, let me
consider the specific issue of measuring the partial width
$\Gamma(H\to\mu^+\mu^-)$. This can be done \cite{gunion2}
combining the results of the {\muc} and TESLA about
$\Gamma(H\to\mu^+\mu^-)\cdot B(H\to b\bar b)$ and $B(H\to b\bar
b)$ respectively. With $L=0.4$ fb$^{-1}$ for the {\muc} and 200
fb$^{-1}$ for TESLA, one can determine this width with a $4\%$
error, implying a $3-\sigma$ sensitivity for $m_A\le 600~GeV$.
This is very important since in this case there is no theory
uncertainty, as for the $b\bar b$ case, coming from possible
decays in SUSY particles, radiative corrections, etc.

{\it SUSY $H^0$ and $A^0$} - The possibility of discovering the
neutral heavy Higgs bosons of the MSSM, $H^0$ and $A^0$, are
limited both at the LHC and the NLC. In the first case since
there is a not accessible region in the plane $(m_A, \tan\beta)$.
In the second case there is a kinematical limit since they should
be pair-produced. If these particles are discovered at the LHC
and/or at the NLC, the {\muc} can make a detailed study of these
particles for any value of $\tan\beta$. Otherwise they could be
discovered through scanning or, if $\tan\beta$ is large enough,
through the bremstrahlung tail \cite{barger}. However, if both
$m_A$ and $\tan\beta$ are large, the $H^0$ and $A^0$ masses get
closer. In this case the {\muc} offers a unique possibility
since, to discriminate between the two resonances, a value of
$\sigma_M$ smaller that the mass difference is required. For
instance, in the case of $m_A=350~GeV$ and $\tan\beta=10$, the two
resonances can be discriminated at the {\muc} with $R=0.06\%$,
whereas this would not be possible at the NLC where the best
value of $R$ is about $0.1\%$.

{\it PNG bosons} - Any theory of dynamical symmetry breaking with
a symmetry group larger than $SU(2)_L\otimes SU(2)_R$ contains
PNGB's. In many models the lightest PNGB, $P^0$, is the colorless,
neutral, $T_3=-1/2$ mass eigenstate \cite{JHEP}. This implies that
$P^0$ couples to muons as the Higgs, proportionally to $m_\mu/v$.
The typical $P^0$ widths in the mass range $50\le m_{P^0}\le
200~GeV$ are $2\le \Gamma_{P^0}\le 20~MeV$. There are no
constraints on $P^0$ from the actual machines, but there are good
possibilities to detect them, if they exist, at the Tevatron Run
II in the case of $m_{P^0}>60~GeV$, and at the LHC in the mass
range $30\le m_{P^0}\le 200~GeV$. On the other hand to detect
$P^0$ at the NLC would be rather difficult, since it has no
tree-level couplings $ZZP^0$, but it is coupled only through a
fermion loop. The main decay mode would be $e^+e^-\to \gamma
P^0$. In practice only the {\muc} has the potential to study this
particle in detail. At the {\muc} one can study the $P^0$ through
the $s$-channel production, but again one needs to use the
smallest possible value for $R$ and to keep errors on $\sigma_M$
at the per cent level. In this case the statistical errors, as
reported in the following Table for a 4 years run with a total
luminosity of $0.4$ fb$^{-1}$, are quite small.

\begin{center}
\begin{tabular}{|c|c|c|c|c|c|c|}
\hline
 Quantity & \multicolumn{6}{c|}{Errors for the scan procedure} \\
\hline \hline
 {\bf Mass (GeV)} & {\bf 60} & {\bf 80} & {\bf $\mz$} & {\bf 110}
& {\bf 150} & {\bf 200} \\
\hline
$\sigma_cB$ &  0.0029 & 0.0054 & 0.043 & 0.0093 & 0.012 & 0.018\\
\hline
 {$\gampzero$} & {0.014} & {0.029} & {0.25} & {0.042}
 & {0.052} & {0.10} \\
 \hline
\end{tabular}
\end{center}
In conclusion a {\muc} offers various advantages with respect to
electron-positron machines, mainly related to the very small
energy spread and to the unique flavor property of the beam.
Still, for the Higgs analysis, a better luminosity of the ones
considered possible so far is desirable. For other cases as the
MSSM heavy Higgs and for the PNG bosons this machine will be able
to allow very precise measurements. I have not considered here
many other possibilities which have been explored in great detail
in refs. \cite{CERN,barger}.

%
\end{document}